\documentclass[pdffig]{article}
\usepackage{graphicx}

\begin{document}
\title{Conditions for the validity of the quantum Langevin equation}
\date{22 September 2011}
\maketitle
\begin{center}

{J. Frenkel}

{jfenkel@fma.if.usp.br}

{Instituto de F\'{i}sica, Universidade de S\~{a}o Paulo, S\~{a}o Paulo,
Brazil }

{J. C. Taylor}

{jct@damtp.cam.ac.uk}

DAMTP, Centre for Mathermatical Sciences, University of Cambridge, Cambridge, UK

\end{center}

\section*{Abstract}
From microscopic models, a Langevin equation can in general  be derived only as an approximation.
Two possible conditions to validate this approximation are studied. One is, for a linear Langevin
equation, that the frequency of the Fourier transform should be close to the natural  frequency of the system. The other is by the assumption
of `slow' variables. We test this method by comparison with an exactly soluble model, and point out its limitations. We base our discussion on two
approaches. The first is a direct, elementary treatment of Senitzky. The second is via
a generalized Langevin equation as an intermediate step.

PACS:  05.10.Gg, 05.30.-d, 05.40.Ca
\section{Introduction}
The Langevin equation is certainly not exact (except in the special case when  the underlying
equations of motion are all  linear), and so it is of interest to find the conditions for it to be a good
approximation.

The equation has been justified, starting from  microscopic equations of motion, by various general arguments and in some
particular models. These arguments start from the Heisenberg equations of motion for a simple (macroscopic
or mesoscopic) system in interaction with an environment (or bath) with many degrees of freedom.
We will assume the systems are quantum ones, but everything we say would equally apply to classical systems (replacing commutators by Poisson brackets, etc.) We choose  a microscopic model
in which the interaction is bilinear in the environment variables. This is the simplest change form the
(trivial) linear case, and is necessary if the environment is fermionic.

We will concentrate on two approaches. The first is due to Senitzky \cite{senitzky}.
It uses only elementary quantum mechanics, but it is not clear for what ranges of parameters the approximations made are good ones. . We argue that one region in which Senitzky's approximations may be justified
is where the frequency is close to the natural frequency of the simple system (which therefore
assumes that system to be linear). Our approach is to regard the Langevin equation as summing
an infinite subset of perturbation theory terms. The question is then, when does this subset
dominate?

For an earlier discussion of Senitzky's argument see \cite{taylor}. So far as we know,
text books do not offer a derivation as simple and direct as Senitzky's. See for example \cite{coffey},
\cite{gardiner}, \cite{kogan}, \cite{balescu}, \cite{kreuzer}, \cite{lebellac}.

The second method of justification is indirect. First a generalized Langevin equation is 
established by a projection method \cite{zwanzig1}, \cite{mori}, \cite{zwanzig}. This is formally exact, but probably not useful. (It is usually restricted to the  case of a linear system, equation (2.3).)
In section (5.2) we point out that in the generalized Langevin equation the separation
into a noise term and a dissipative term is ambiguous.

Then it is assumed that the system variables are `slow' compared to the environment ones.
It is argued that the generalized Langevin equation then simplifies greatly, and becomes
an ordinary Langevin equation.
In section 5, we examine this proposal critically, and test it in an exactly soluble  example. For some text book accounts, see
\cite{lebellac}, \cite{zwanzig}, \cite{coffey}.

It is worth noting that both the above methods work by manipulating the underlying
Heisenberg equation of motion. Yet the Langevin equation in the end refers to
an expectation value over some distribution function, chosen on physical grounds.
Clearly some assumptions have come in during the course of the derivations.

\section{The underlying microscopic systems}

 We use the Heisenberg picture throughout.
 
 We are concerned with attempts to deduce a Langevin equation from an underlying dynamics.
 This consists of a simple macroscopic or mesoscopic system interacting with an environment (sometimes called a bath). For simplicity, we take the system to have one degree of freedom,
 with phase space $Q,P$. The environment has many (microscopic) degrees of freedom
 $q_i,p_i$ with $i=1,...,N$. The Hamiltonian is
 $$H=H_0(t)+H'(t),\,\,\,\,\,H_0(t)=H_S(t)+H_E(t)\eqno(2.1)$$
 $$H_S=(1/2)P^2/M+V(Q),\,\,\,\,H_E=\sum_{i=1}^N(1/2) (p_i^2/m_i +m_i\omega_i^2q_i^2),\,\,\,H'=-\alpha QK(q_i,p_i)\eqno(2.2)$$
 ($N$ large).
 Thus we are assuming the environment to be a set of oscillators. We will sometimes
 take
 $$V(Q)=(M/2)\Omega^2Q^2 \eqno(2.3)$$
 that is the system is an oscillator too. In this case, the resulting Langevin equation is linear.
  In (2.2), $\alpha$ designates the coupling strength. We have assumed that $H'$ is linear
  in the system variables, and for simplicity chosen $Q$ (not $P$) to appear there.
  
  It will be useful to define operators $Q_0,P_0,q_{0i},p_{0i}$ to coincide with $Q,P,q_i,p_i$
  at an initial time $t_0$, but to vary with time according to the free ($\alpha=0$) equations of
  motion. That is
  \pagebreak
  $$Q_0(t_0)=Q(t_0).\,\,\,P_0(t_0)=P(t_0),$$
  $$ \hbar \dot{Q}_0(t)=i[H_{S0},Q_0(t)],\,\,\,\hbar \dot{P}_0(t)=i[H_{S0},P_0(t)],$$
  $$H_{S0}=H_S(Q_0(t),P_0(t)),$$
  $$q_{0i}(t_0)=q_i(t_0),\,\,\, p_{0i}(t_0)=p_i(t_0),$$
  $$\hbar\dot{q}_{0i}(t)=i[H_{E0},q_{0i}(t)],\,\,\,\hbar\dot{p}_{0i}(t)=i[H_{E0},p_{0i}(t)],$$
  $$H_{E0}=H_E(q_{0i}(t),p_{0i}(t)).\eqno(2.4)$$
  Note that $H_{S0}$ and $H_{E0}$ are each independent of $t$,

  If the function $K$ in (2.2) is linear, the equations of motion of the environment are linear,
  and these variables can be simply eliminated to produce an exact Langevin equation.
  No statistical distribution function appears in this Langevin equation. For these two reasons,
  the case of $K$ being linear is misleadingly simple, although it is treated in many text books,
  for example as a model of Brownian motion and of decoherence.
  
  The simplest  non-trivial example is for $K$ to be bilinear in the $q_i,p_i$,
  and this is what we shall assume below. Note that the environment might consist
  of fermionic variables (like conduction electrons), and then $K$ would necessarily
  be bilinear or of higher degree. We will not treat this fermionic case explicitly,
  but our arguments below can easily by generalized to cover fermi statistics.
  
  The remaining  ingredient is a statistical distribution function. In non-equlibrium statistical
  physics, the choice of distribution function is a matter of physical judgement.
  We shall mention just two possibilities. The first is simply
  $$\rho=Z^{-1}\exp(-H/T) \eqno(2.5)$$
  the equilibrium distribution. The second is the factorized free distribution
  $$\rho_0=Z_0^{-1}\exp(-H_0/T)=\rho_S\times\rho_E   =\rho_S \times Z_E^{-1}\exp(-H_{E0}/T)\eqno(2.6)$$
   where $H_{S0}$ is the time-independent energy defined in (2.4), and we need not specify $\rho_E$
   further.
   Units of temperature $T$ are chosen so that Boltzmann's constant is unity,
  and the partition functions $Z$, $Z_0$ and $Z_E$ are normalization factors. The use of $\rho_0$ is motivated by the idea that the environment is initially
  at equilibrium by itself, and then the system is brought into contact with it at some initial time $t_0$.
  Since we use the Heisenberg picture, density matrices are time-indpendent, and so the
  factorization property (2.6) is a single condition, not one for each value of $t$ as it would be
  in the Schrodinger picture. But $H_{E0}$, and therefore also $\rho_E$, depend implicitly on the
  initial time $t_0$.
  For the Hamiltonian $H_E$ in (2.2), $\rho_{E0}$ factorizes further:
  $$\rho_{E0}=\prod_i \rho_{i0} \eqno(2.7)$$
  
  We define expectation values, for an operator X,
  $$\langle X\rangle=\mathrm{tr}(X\rho),\,\,\,\,\langle X\rangle_0=\mathrm{tr}(X\rho_0).\eqno(2.8)$$

In its simplest form, the Langevin equation  which one might hope to derive, is usually assumed to be (for $t>t_0$)
$$M\ddot{Q}(t)+V'(Q(t))+\alpha^2 \int_{t_0}^{\infty} dt' \bar{C}(t-t')Q(t')=\alpha K_0(t),\eqno(2.9)$$
where
$$\bar{C}(t,t')=\langle C(t,t')\rangle_0=\langle C(t,t')\rangle_E\equiv \mathrm{tr}\{C_E(t,t')\rho_E \}\eqno(2.10)$$
and
$$C(t,t')=\theta(t-t')c(t,t'),\,\,\,\,c(t,t')=-{i\over\hbar }[K_0(t), K_0(t')] =C(t,t')-C(t',t)\eqno(2.11)$$
and  $K_0=K(q_{0i},p_{0i})$.
(In the classical case, the right hand side  of (2.11) is to be replaced by the  Poisson bracket 
and the trace in (2.8) by $\int\prod_i dq_idp_i$.) Equation (2.9) shows the characteristic features
of a Langevin equation: the noise $K_0$ and dissipation contained in the $C$ term (non-Markovian in general,
 that is  frequency dependent.)

If $K$  is linear in the environment variables, (2.9) is exact .This follows simply by solving for
the $q_i(t)$ (the retarded solution) in terms of $Q$ and $q_i(t_0),p_i(t_0)$ and inserting
this solution into $K(q_i,p_i)$.
 In this special case $C(t,t')$ is a $c$-number, and in fact is not dependent of any dynamical variables;
 so the expectation value in (2.10) is redundant.

A typical quantity one might want to compute by using (2.9) is the correlation function
$$ S_Q(t,t')=\langle Q(t)Q(t')\rangle_0\equiv \mathrm{tr}\{ Q(t)Q(t')\rho_0\}.\eqno(2.12)$$
In general, $S_Q$ may not be a function of $(t-t')$ only. This is because $\rho_0$ defined in (2.7)
does not commute with $H'$ and so not with the total Hamiltonian $H$.
If $t$ and $t'$ are much later than $t_0$, the form of $\rho_S$ may be unimportant in (2.12).

 There is a general result, the fluctuation-dissipation theorem.  Define
 $$S_{K_0}(t-t')=\langle K_0(t)K_0(t')\rangle_E\equiv \mathrm{tr}\{K_0(t)K_0(t')\rho_E\} \eqno(2.13)$$
 (note that $S_K$  does depend only on the time difference because $\hbar\dot{K}_0=i[H_E,K_0]$ and $[H_E,\rho_E]=0$) and define the Fourier transform by
 $$S_{K0}=\int d\omega e^{-i\omega(t-t')}\tilde{S}_K(\omega) .\eqno(2.14)$$
 Then the relation is
 $$-i\hbar\tilde{\bar{c}}(\omega)=\tanh\left({\hbar\omega \over 2T}\right)\left(\tilde{S}_K(\omega)+\tilde{S}_K(-\omega)\right)\eqno(2.15)$$
 where $\tilde{\bar{c}}$ is the Fourier transform of the expectation vlue of $c$ in (1.8).
 
  In section 3 we review Senitzky's derivation of a Langevin equation, and then
 derive one condition for its validity. Section 4 reviews the so-called generalized Langevin
 equation, and section 5 discusses how the ordinary Langevn equation might follow if
 $Q,P$ are `slow' variables. Section 6 summarizes our conclusions.

\section{Senitzky's argument}

We first emphasize the salient features of Senitzky's \cite{senitzky} argument, which is quite general.

 By using the Heisenberg equations of motion,
for $K$ and $H_E$, Senitzky derives the exact equations
$$M\ddot{Q}+V'(Q)=\alpha K, \eqno(3.1)$$
$$K(t)=K_0(t)-{i\alpha\over \hbar}\int_{t_0}^t dt_1\int_{t_0}^{t_1}dt_2U^*(t-t_1)[K(t_1), \dot{K}(t_2)Q(t_2)]U(t-t_1), \eqno(3.2)$$
where 
$$i\hbar\dot{K}(t)=[K(t),H]\equiv [K(t),H_E(t)+H_S(t)+H'(t)]=[K(t),H_E(t)], \eqno(3.3)$$
 
$$U(t)=\exp(-itH_E(t_0)/\hbar)=\exp(-itH_{E0}/\hbar).\eqno(3.4)$$
(Note that it is $H_E(t)$ which comes in (3.3), not $H_{E0}$.)

Senitzky then approximates in (3.2) $K$ by $K_0$  and $U^*(t-t_1)Q(t_2)U(t-t_1)$ by $Q(t_2)$, to obtain
$$K(t)\simeq K_0(t)-{i\alpha\over \hbar}\int_{t_0}^t dt_1\int_{t_0}^{t_1}dt_2[K_0(t),\dot{K}_0(t_2+t-t_1)]Q(t_2).\eqno(3.5)$$
Now the $t_1$ integration may be done to give
$$K(t)\simeq K_0(t)-\alpha\int_{t_0}^t dt' C(t,t')Q(t'), \eqno(3.6)$$
where $C$ is given by (2.11). 

Equation (3.6) is the first stage of Senitzky's approximation.
The second is to use to use the further approximation
$$ C(t,t')\simeq\bar{C}(t,t'), \eqno(3.7)$$
giving, together with (3.1), the Langevin equation (2.6).
(Note that the approximation (3.7)  is needed in the classical case as well as the quantum one.
Only in the linear case is (2.9) exact.)

The approximations leading to (3.6)
amounts to neglecting in $K$ some, but not all, terms down by a power of $\alpha^2$. 
In order to discuss this, we define
the power series
$$K=K_0+\alpha K_1+\alpha^2 K_2+....,\,\,\,Q=Q_0+\alpha Q_1+\alpha^2 Q_2+....\eqno(3.8).$$
Then for example a neglected term containing $K_1Q_0$ is not obviously smaller than the retained
term $K_0Q_1$ in (3.5). Thus it is far from obvious that (3.5) is a valid approximation.

\subsection{Perturbation theory}

In order to investigate the region of validity of Senitzky's first approximation (3.5), we
consider an expansion of $K$ in powers of $\alpha$. The approximation picks
out an infinite subset of terms. The question is, when do these terms dominate.
We shall attempt to answer this question by looking at the lowest order terms, but we believe that
our argument generalizes to all orders.

 The solution of the exact
equation (3.4) is more easily derived direcly from the original Heisenberg equations of motion
 (3.3) etc. It is
$$K(t)=W^*(t)K_0(t)W(t), \eqno(3.1.1)$$
where $W$ is the solution of
$$ i\hbar\dot{W}(t)=H'_0(t)W(t)=-\alpha K_0(t)Q_0(t)W(t),\;\;\;W(t_0)=1. \eqno(3.1.2)$$

We may now compare the order $\alpha^2$ term, $K_2$, from (3.1.1) with the approximation from (3.5).
To order $\alpha^2$, the exact (3.1.1) gives
$$K_2(t)=-{1\over \hbar^2}\int_{t_0}^t dt_1\int _{t_0}^{t_1} dt_2[[K_0(t),K_0(t_1)Q_0(t_1)],K_0(t_2)Q_0(t_2)]. \eqno(3.1.3)$$

 In order to find the approximate form of $K_2$ deduced from (3.5), we first find $Q_1$ . This is derived (3.1),
 with $K$ on the right approximated by $K_0$.   The solution of this  equation may be written
 $$Q_1(t)={i\over \hbar}\int_{t_0}^t dt'[Q_0(t),Q_0(t')]K_0(t') .\eqno(3.1.4)$$
 (This is easy to check if $V$ is a sum of integral powers of $Q$.)
 
 Inserting (3.1.4) into (3.6) gives (using (2.11))
 $$K_2(t)\simeq -{1\over \hbar^2}\int_{t_0}^t dt_1\int _{t_0}^{t_1} dt_2[K_0(t),K_0(t_1)]K_0(t_2)[Q_0(t_1),Q_0(t_2)]. \eqno(3.1.5)$$
 The difference between (3.1.3) and (3.1.5) is
 $$ -{1\over \hbar^2}\int_{t_0}^t dt_1\int _{t_0}^{t_1} dt_2[[K_0(t),K_0(t_1)],K_0(t_2)]Q_0(t_2)Q_0(t_1). \eqno(3.1.6)$$

 In general, (3.1.5) and (3.1.6) are  of the same order. It is clear that the Langevin
 equation sums a subset of the terms in (3.1.1). The question is: what are the conditions for
 this subset to dominate over the other terms, like (3.1.6)?  We are able to propose an answer to this
 question, but only for the case that the system is an oscillator, that is assuming (2.3). Then the free equation for $Q(t)$ has a Greens function
 $$G(t)= \theta(t) (M\Omega)^{-1}\sin(\Omega t). \eqno(3.1.7)$$
 and
 $$\theta(t-t')[Q_0(t),Q_0(t')]=-i\hbar G(t-t'). \eqno(3.1.8)$$
 The Fourier transform of (3.1.7) is
 $$\tilde{G}(\omega)=-{1\over (2\pi M)}{1\over (\omega+i\epsilon)^2-\Omega^2}.\eqno(3.1.9)$$
 The structure of (3.1.5), but not of (3.1.6), allows for a pole in (3.1.9) at $\omega=\Omega$
 to appear in suitable functions of $\omega$. But to complete this argument, we must 
 include Senitzky's second approximation, and to do this we need a more detailed model
 for $K$, which is the subject of the next subsection.

 \subsection{A model interaction}
In this subsection we will take $t_0=-\infty$. 
We discuss later what changes if $t_0$ is finite.

In order to test Senitzk's argument, we take the simplest model we can think of, which
is not totally linear. We take $K$ in (2.2) to be bilinear in the environment variables.
If we have in mind a model for Brownian motion, then the simple case when $K$ is linear
might be thought of as representing the Brownian particle emitting phonons.
In our bilinear generalization, phonons scatter off the particle. Of course, these are only crude
models.

Our formalism can easily be extended to a fermionic environment, when $K$ cannot be linear
but can be bilinear.

Define annihilation operators by
$$a_i(t)={ip_i(t)+m_i\omega_iq_i(t)\over \sqrt{2m_i\omega_i}},\,\,\,\,\,a_i(t)=\exp(-i\omega_it)a_i.\eqno(3.2.1)$$
Then we take
$$  K=\hat{K}-\langle\hat{K}\rangle_E\,\,\, \mathrm{where}\,\,\,\hat{K}(t)=L^{-1}\sum_{i,j=1}^N(\omega_i \omega_j)^{1/2}a^*_i(t)b_{ij}a_j(t), \eqno(3.2.2)$$
where
 the $b_{ij}$ are dimensionless numbers and $L$ is a parameter with dimensions of length
which we shall not specify further (but which might represent for example a lattice spacing).
With $K$ defined as in (3.2.2), $\alpha$ in (2.2) is dimensionless.

We shall concentrate on $c$, the odd part of $C$ in (2.11).
From (3.2.2),
$$c(t,t')=\sum_{ij}(\omega_i\omega_j)^{1/2}X_{ij}(t,t')a_i^*a_j \eqno(3.2.3)$$
where
$$X_{ij}(t,t')=-iL^{-2}\sum_k  \omega_k b_{ik}b_{kj}[\exp\{i(\omega_i-\omega_k)t+i(\omega_k-\omega_j)t'\}-(i\leftrightarrow j)].\eqno(3.2.4)$$

From (3.2.3), (3.2.4),
$$\bar{c}(t-t')\equiv \langle c(t,t')\rangle_E=\sum_i X_{ii}w(\omega_i)\eqno(3.2.5)$$
where
$$X_{ii}(t-t')=-2L^{-2}\sum_{k}\omega_k b_{ik}b_{ki}\sin[(\omega_k-\omega_i)(t-t')] \eqno(3.2.6)$$
and
$$ w(\omega)=\hbar\omega
 [\exp(\hbar\omega/T)-1]^{-1}. \eqno(3.2.7)$$
The classical (high temperature) limit of $w(\omega)$ is $T$, and the low temperature limit is $\hbar \omega\exp(-\hbar\omega/T)$.

We first discuss the Senitzky's second approximation (3.7), and return to (3.5) later. So we begin by assuming (3.6)
and combine it with (3.1) to give
$$M\ddot{Q}+M\Omega^2 Q+{\alpha^2\over 2}\int_{-\infty}^tdt'\{C(t,t')Q(t')+Q(t')C(t,t')\}=\alpha K_0(t) \eqno(3.2.8)$$
 (We have symmetrized the
order of the operators in the integrand in order to make it explicitly Hermtian. Since $[C,Q]=O(\alpha)$ this involves only higher
order terms in the integrand.) 

We will formally
solve (3.2.8) as a power series in $\alpha$, and then compare the terms in this series, with and without
the use of (3.7). Of course, the Langevin equation sums an infinite number of orders of
$\alpha$. but approximations which fail term-by-term in the power series are unlikely to succeed for the complete series.

In order to be definite, we assume we are using the Langevin equation in order to calculate the $Q$-noise correlation function (2.12). It is this assumption which gives a physical motivation for using $\rho_0$ as the distribution function. Up to this point, there was no reason to prefer $\rho_0$ in the 
 approximation (3.7).

Through order $\alpha^4$, we get  
$$S_Q(t-t')=\langle Q_0(t)Q_0(t')\rangle_0+\alpha^2\int dt_1dt_2G(t-t_1)G(t'-t_2)\langle K_0(t_1)K_0(t_2)\rangle_0$$
$$+{\alpha^4\over 2}\int dt_1dt_2dt_3t_4G(t'-t_4)G(t-t_1)G(t_2-t_3)$$
$$\times\langle \{C(t_1,t_2)K_0(t_3)+K_0(t_3)C(t_1,t_2)\}K_0(t_4)\rangle_0$$
$$ +(\mathrm{herm.} \, \mathrm{conj.} \,\,\mathrm{with} \, t\leftrightarrow t')\eqno(3.2.9)$$
(Note that, using $t_0=-\infty$, $S_Q$ turns out to depend only on $t-t'$, athough this is not obvious from
the definition (2.12).)
If the approximation (2.7) were valid, we should be able to approximate the expectation value
at the end of (3.2.9) by
$$\langle C(t_1,t_2)K_0(t_3)K_0(t_4)\rangle\simeq
 \bar{C}(t_1-t_2)\langle K_0(t_3)K_0(t_4)\rangle,\eqno(3.2.10)$$
etc. 
So this is what we now check, using (3.2.3) and (3.2.4).

To evaluate the expectation value in (3.2.10), we need (using (3.2.2))
$$\sum_{ijklmn}X_{ij}b_{kl}b_{mn}\mathrm{tr}\left(a^*_ia_ja^*_ka_la^*_ma_n\prod_r \rho_{r0}\right). \eqno(3.2.11)$$
This receives nonzero contributions from the following values
$$\mathrm{(a)}:\,i=j,\,k=n,\,l=m;\,\,\,\mathrm{(b)}:\,i=l,\,k=n,\,j=m;\,\,\,\mathrm{(c)}:\,i=n,\,j=k,\,l=m\eqno(3.2.12)$$
where in each case we assume there are no other equalities. There are also contributions like
$i=j=k=n,\,l=m$, but for large $N$ these are negligible compared to (3.2.12).

The contribution to (3.2.11) from region (a) in (3.2.12) gives the right hand side of (3.2.10); so
the question is, when are the contributions from (b) and (c) small compared to that from (a)?

 The contribution to $\tilde{\tilde{C}}$ from (3.2.9) has the form
 $$i\alpha^4[\tilde{G}(\omega)]^2\sum_{ijkl}[W_a+W_b+W_c]+\mathrm{herm. conj.} \eqno(3.2.13)$$
 where $W_a,W_b,W_c$ come from the ranges in (3.2.12):
 $$W_a=\tilde{G}(\omega)w(\omega_j)w(\omega_l)b_{kl}b_{lk}b_{ij}b_{ji}\{(\omega-\omega_i+\omega_j+i\epsilon)^{-1}\delta(\omega-\omega_k+\omega_l)$$
$$-(\omega_{i,j,k,l}\rightarrow -\omega_{i,j,k,l}), \eqno(3.2.14a)$$
$$W_b=\tilde{G}(\omega_i-\omega_k)w(\omega_j)w(\omega_l)b_{ij}b_{jl}b_{lk}b_{ki}(\omega-\omega_i+\omega_j+i\epsilon)^{-1}\delta(\omega-\omega_l+\omega_j), $$
$$-(\omega_{i,j,k,l}\rightarrow -\omega_{i,j,k,l}), \eqno(3.2.14b)$$
$$W_c=\tilde{G}(\omega_j-\omega_l)w(\omega_j)w(\omega_l)b_{ij}b_{jl}b_{lk}b_{ki}[(\omega-\omega_i+\omega_j+i\epsilon)^{-1}\delta(\omega-\omega_k+\omega_l))$$
$$-(\omega_{i,j,k,l}\rightarrow -\omega_{i,j,k,l}), \eqno(3.2.14c)$$
Here $\tilde{G}$ is defined in (3.1.9), and has a pole where its argument is equal to the natural
frequency $\Omega$ .Tthere are three
such poles at $\omega=\Omega $ in (3.2.13) from(3.2.14a), whereas from the other two terms there are only two poles.    
Thus $W_a$, and so  the approximation (3.2.10) to (3.2.9), may be good for values 
of $\omega$ sufficiently near to $\Omega$. Otherwise,we can see no reason why $W_a,W_b, W_c$
should not be comparable.

The structure of the terms in equations (3.2.14) is illustrated in Fig. 1. The generalization to all orders
of $\alpha$ is that the Langevin equation sums `bubble graphs', that is graphs
like the first in the figure, with a sequence of `bubbles' like the one marked $X$ in the figure,
connected by $Q$-propagators. These graphs have the maximum number of poles at
$\omega=\Omega$.
\begin{figure}
\centering
\includegraphics[scale=0.7]{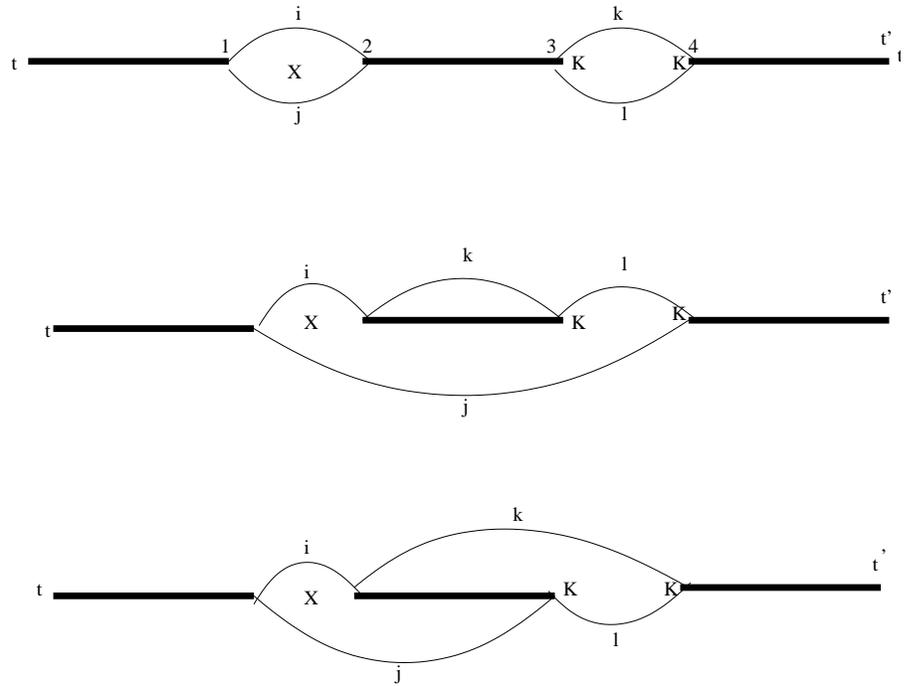}

\caption{Graphs symbolizing the terms in (3.2.14), Thick lines represent the $Q$-propagator (3.1.9).
Thin lines indicate the pairings in (3.2.12). The $X$ represents the commutator $c$ in (2.11).
and black circles represent the $K_0$ operators in (3.2.9). The vertices are labeled according to
the values of the times in (3.2.9).}
\end{figure}

We now turn to the first approximations made in Senitzky's argument, the replacement of (3.2) by (3.5).
To order $\alpha^2$, this implied the neglect of (3.1.6) compared to (3.1.5).  If we work out the contribution of (3.1.5) to (3.2.9), we need instead of (3.2.10):
$$\langle [C(t_1,t_2),K_0(t_3)]K_0(t_4)\rangle_0. \eqno(3.2.15)$$
 A contribution like  (3.2.14a) it is zero, because $[\tilde{C}(t_1,t_2),K_0(t_3)]=0$.
Thus (3.1.5) has parts which with three poles at $\omega=\Omega$, but (3.1.6) does not,
so near this pole both Senitzky's approximations (3.5) and (3.7) are justified.

If $t_0$ is finite, there are no poles at $\omega=\Omega$. However, in the Fourier transform of (3.2.9),
times less than $t_0$ do not contribute much if
 $$\omega |t_0|\gg 1 \eqno(3.2.16).$$
 Thus we expect the would-be poles at $\omega=\Omega$ to be large provided that 
 $|t_0|$ is large enough to satisfy (3.2.16).

\section{Non-linear $Q$-systems}

Most of our arguments above have assumed that the simple system is an oscillator, as in (2.3), with
a well-defined natural frequency $\Omega$, so that we can recognize the poles (3.1.9).
But we may ask if the arguments extend to the more general potential $V(Q)$ in (2.2), when the
equation of motion for $Q$ is nonlinear. Then a simple Greens function like (3.1.9) does not
exist. However, when working to first order,  there is the equation (3.1.4),
where $\theta(t-t')[Q_0(t),Q_0(t')]$ looks a bit like a Greens function, though in general it is an
operator and is a function of $t$ and $t'$ separately.

Let us consider an example like (3.2.9). For a non-linear system, each Greens' function $G(t-t')$ is
replaced by the operator $i\theta(t-t')[Q_0(t),Q_0(t')]/\hbar$. We have to take the expectation value 
of products of these commutators using $\rho_S$ in (2.6). In order to make the same sort of
argument as we did with the propagator poles (3.1.9), two things are needed: the expectation
values of products commutators should approximately factorize into products
of expectation values $\langle Q_0(t)Q_0(t') \rangle$; and the Fourier transforms of each of these should
have a pole at some definite frequency. Whether or not these things happen in any useful
approximation will depend upon the potential $V(Q)$.

As the simplest example, take a product of two commutators, as would occur in the first line of (3.2.9).
In the nonlinear 
case, the product of two Greens functions $G(t-t_1)G(t'-t_2)$ would be replaced by
$$-(1/\hbar^2)\theta(t-t_1)\theta(t'-t_2) \mathrm{tr}\{[Q_0(t),Q_0(t_1)][Q_0(t'),Q_0(t_2)]\rho_S\} ,\eqno(4.1)$$
Let us assume that $H_S$ has a discrete set of energy eigen-states $|\alpha\rangle$
with energy $E_{\alpha}$. and take the simplest case where $\rho_S$ corresponds to the (pure)
ground state $|0\rangle$.
Then (4.1) has contributions
$$-(1/\hbar^2)\theta(t-t_1)\theta(t'-t_2)\sum_{\alpha,\beta,\gamma}\Big(\langle 0|Q_0(t)|\alpha\rangle\langle\alpha|Q_0(t_1)|\beta\rangle 
\langle \beta|Q_0(t')| \gamma\rangle\langle\gamma|Q_0(t_2)|0\rangle$$
$$-(t\leftrightarrow  t_1)-(t'\leftrightarrow t_2)+(t,t'\leftrightarrow t_1,t_2)\Big).\eqno(4.2)$$
We will get the required behaviour if there is a significant contribution to (4.2) from $|\beta\rangle =|0\rangle$ and $|\alpha\rangle=|\gamma\rangle$ for some $|\alpha\rangle$.
If this happens, the time-dependence of (4.2) is
$$\theta(t-t_1)\theta(t'-t_2)\sin\{(t-t_1)(E_{\alpha}-E_0)/\hbar\} \sin\{(t'-t_2)(E_{\alpha}-E_0)/\hbar\} ,\eqno(4.3)$$
and then the Fourier transform has  poles at the frequency
$$\Omega_{0\alpha}\equiv (E_{\alpha}-E_0)/\hbar \eqno(4.4)$$
which looks like what one gets from (3.1.7).
But of course, as well as (4.3), there will be other contributions to  (4.2), which are not
functions of just the two variables $(t-t_1)$ and $(t'-t_2)$. So there is in general no reason to expect
(4.4) to dominate.

An heuristic quantum Langevin equation  has been used for a system involving a shunted Josephson junction \cite{coffey}, \cite{koch}, \cite{beck}, \cite{brandt}, \cite{levinson}. Then  $Q$ is the Josephson angle
and
$$V(Q)=-IQ-I_0\cos(Q) \eqno(4.5)$$
 with the constant coefficients satisfying $I>I_0$. In this potential,  there are no discrete quantum states, so the
 above considerations are irrelevant. What  is more, in the classical
 motion $\dot{Q}$ increases indefinitely;
but, when the damping due to the Langevin equation is included, the motion is in general qualitatively very different.
So it is not possible to use an expansion in powers of the interaction strength $\alpha$
(except perhaps for a short time interval after the initial time $t_0$).  So in this case we are not able
to study the validity of the approximations leading to the Langevin equation.

Thus we are unable to find any conditions which would validate Senitzky's approximations
for a general potential $V(Q)$.

\section{Generalized Langevin equations}

Another route to a Langevin equation is via a generalized Langevin equation, constructed with the aid
of projection operators, followed by the assumption that $Q,P$ are "slow" variables (see for example 
 \cite{lebellac}, \cite{zwanzig}, \cite{zwanzig1}, \cite{mori}, \cite{grabert}). 
 
 \subsection{Projection operators}
 
 In the dynamics defined in (2.1) and (2.2), we will assume that $Q,P$ are `slow"'(or `relevant') and $q_i,p_i$
 are `fast' variables. We need a projection operation, projecting onto the `slow' subspace.
 This entails a scalar product between operators, which we will choose to define as (for Hermitean
 operators $A$ and $B$)
 $$(A,B)_0=(1/2)\langle (AB+BA)\rangle_0 \equiv (1/2)\mathrm{tr}\{AB+BA)\rho_0\}. \eqno(5.1.1)$$
 This is symmetric which is convenient. (Some authors \cite{lebellac} use the more complicated  Mori product.
 For simplicity, we do not take this course.)
 
  We have defined (5.1.1) using $\rho_0$, defined in (2.6). A disadvantage of (5.1.1)
 is that
 $$(A,[H,B])_0\neq -([H,A],B)_0. \eqno(5.1.2)$$
 Another possibility, which does not have this disadvantage, is to use the equilibrium distribution $\rho$.
 We will adopt this second choice in section 5.3 below, and then drop the suffices 0 in (5.1.1)
 
 From now on, for simplicity, we will set $t_0=0$.
 
 Given (5.1.1), the projection operation onto the "slow" subspace is defined to be, for any operator $X$,
 $${\mathcal{P}}_0X={(X,Q(0))_0\over (Q(0),Q(0))_0}Q(0)+{(X,P(0))_0\over (P(0),P(0))_0}P(0), \eqno(5.1.3)$$
 and the complementary projection is $(1-\mathcal{P}_0)$. It seems to be usual to define
 $\mathcal{P}_0$ in terms of $Q,P$ at the initial time, and we have emphasized this in (5.1.3).
 In general, a different choice of time would define a different projector. Note that
 $(Q(0),P(0))=0$ because of invariance under time-reversal.
 
 Obviously
 $${\mathcal{P}}_0Q^2(0)\neq Q(0), \eqno(5.1.4)$$
 and for this reason the method may not be appropriate except in the linear case (2.3) (see section 8.2
 of \cite{zwanzig}). Also
 $${\mathcal{P}}_0Q(t)\neq Q(t)\,\,\,\,(t\neq 0). \eqno(5.1.5)$$
 (For this reason, the projection method may be more appropriate in the Schr\"{o}dinger picture,
 as in the derivation of the master equation.)
 But, for the linear case, we do have
 $${\mathcal{P}}_0Q_0(t)= Q_0(t), \eqno(5.1.6)$$ 
 and 
 $${\mathcal{P}}_0g(q_{0i}(t),p_{0i}(t))=0,\,\,\,\,{\mathcal{P}}_0\{Q_0(t)g(q_{0i}(t),p_{0i}(t)\}=Q_0(t)\langle g(q_{0i}(t),p_{0i}(t)\rangle_0 \eqno(5.1.7)$$
 etc., where $g$ is any function. These equations would not be true if we had defined $\mathcal{P}$ in terms of $\rho$ rather than $\mathcal{P}_0$ in terms of  $\rho_0$.
 
 \subsection{Generalized Langevin Equation}
 
 We need also to define the Liouville operator $\mathcal{L}$, acting on an operator $X$,
 by
 $${\mathcal{L}} X=i[H,X]. \eqno(5.2.1)$$
 From the Heisenberg equations of motion, the generalized Langevin equation is deduced (see for example \cite{zwanzig}):
 $$M\dot{Q}(t)=P(t) \eqno(5.2.2)$$
 $$ \dot{P}(t)+M\Omega^2Q(t)-\alpha \Lambda_{KQ}Q(t)-\alpha \Lambda_{KP}P(t)$$
 $$+\int_0^t dt'\{J_Q(t-t')Q(t')+J_P(t-t')P(t')\}
 =F(t), \eqno(5.2.3)$$
 where
 $$F(t)=\exp\{t(1-{\mathcal{P}}_0){\mathcal{L}}\}(1-{\mathcal{P}}_0){\mathcal{L}}P(0) \eqno(5.2.4)$$
 with the property that
 $${\mathcal{P}}_0F(t)=0, \eqno(5.2.5)$$
 $$\Lambda_{KQ}\equiv (K(0),Q(0))_0/(Q(0),Q(0))_0,\,\,\,\,\Lambda_{KP}\equiv (K(0),P(0))_0/(P(0),P(0))_0,\eqno(5.2.6)$$
 and
 $$J_Q(t)=-({\mathcal{L}}F(t),Q(0))_0/(Q(0),Q(0))_0,\,\,\,J_P(t)=-({\mathcal{L}}F(t),P(0))_0/(P(0),P(0))_0.\eqno(5.2.7)$$
 The quantities in (5.2.6) are in fact zero as we are using $\rho_0$ in this section; but
 the corresponding quantities in section 5.3 (using $\rho$) are not both zero,
 
 One should note that (5.2.3) is an exact consequence of the Heisenberg equations of motion,
 and yet it contains reference to the choice of $\rho_0$ in (5.1.1). Any apparent
 dependence on $\rho_0$ must cancel between the terms in (5.2.3). This, of course, may no longer
 be true if any approximations are made to (5.2.3).
 
 In (5.2.3), it is usual to think of $F$ as being some generalized noise, and the $J$ term as 
 representing some (in general non-Markovian) friction. For this interpretation, we would
 hope that
 $$\langle F(t) \rangle_0=0. \eqno(5.2.8)$$ 
 We cannot prove this in general, or even that $\langle F(t)\rangle_0$ is time independent.    (But in this connection, see section 8.3 of \cite{zwanzig}.)
 
 It may throw some light on (5.2.3) to expand $F$ and $J$ through second order
 in $\alpha$. Let
 $$F(t)=\alpha F_1(t)+\alpha^2F_2(t)+.... \eqno(5.2.9)$$
 In (5.2.4),
 $$(1-{\mathcal{P}}_0){\mathcal{L}}P(0)=(1-{\mathcal{P}}_0)\{-M\Omega^2Q(0)+\alpha K(0)\}$$
 $$=(1-{\mathcal{P}}_0)K_0(0)+O(\alpha^2)=K_0(0)+O(\alpha^2), \eqno(5.2.10)$$
 and so, from (5.2.4) and (5.2.7),  to first order
 $$ F_1(t)=\exp\{t(1-{\mathcal{P}}_0){\mathcal{L}}\}K_0(0).\eqno(5.2.11)$$
 Since we are working to first order in $\alpha$ here, we can neglect $H'$ in $\mathcal{L}$, and get
 $$(1-{\mathcal{P}}_0){\mathcal{L}}K_0(0)\simeq(1-{\mathcal{P}}_0)\dot{K}_0(0) =\dot{K}_0(0).\eqno(5.2.12)$$
 (using (5.1.7)),
 and similarly for higher derivatives. Therefore
 $$F_1(t)=\exp(t{\mathcal{L}})K_0(0)+O(\alpha^2)=K_0(t)+O(\alpha^2).\eqno(5.2.13)$$
Inserting (5.2.13) into (5.2.7) we find a contribution (using (2.2))
$$J_Q(t)\simeq -(i/\hbar)\alpha ([H',K_0(t)],Q)_0/(Q,Q)_0=-(i/\hbar)\alpha^2\langle[K_0(t),K_0(0)]\rangle_0, $$
$$J_P(t)\simeq0, \eqno(5.2.14)$$
which, working through second order, is all we need.

We can now infer $F_2$ in (5.2.9). The generalized Langevin equation (5.2.3) is exact,
and our expansion through second order in $\alpha$ must be exact to that order.
But the Langevin equation  (3.1) with (3.6) is also exact through to second order (because the terms neglected in going from (3.1.3) to (3.1.4) were higher than the second). So (5.2.3) must be the same as (3.6)
when (5.2.13) and (5.2.14) are inserted. This requires that
$$F_2(t)=(i/\hbar)\int_0^t dt'\{[K_0(t),K_0(t')]-\langle [K_0(t),K_0(t')]\rangle_0\}Q_0(t').
\eqno(5.2.15)$$
We may check that this satisfies (5.2.5).

Thus we see that the `noise' term (5.2.15) corrects the approximate friction term (5.2.14)
so as to give the correct (through order $\alpha^2$) friction term in (3.6). The interpretation of
$F$ as noise may be open to question.

\subsection{The slow variable approximation}

It has been proposed to derive an ordinary Langevin Equation from the generalized one (5.2.3)
as an approximation assuming that, in our example, $Q$ and $P$ are `slow' variables.
We will express this assumption in the form
$$\dot{Q}=O(\Omega)\times Q, \,\,\,\,\,\dot{P}=O(\Omega)\times P \eqno(5.3.1)$$
where $\Omega$ is small. Presumably this means that $\Omega\ll \bar{\omega}$, where
$\bar{\omega}$ is some sort of typical value of the $\omega_i$ in (2.2),

We will follow the argument as presented in section 8.6 of \cite{zwanzig}. First, we must
depart from the choice $\rho_0$, which we have made up to now, and use the equilibrium
distribution $\rho$ as in \cite{zwanzig}. Then, since $[H,\rho]=0$,
$$(X,{\mathcal{L}}Y)=-({\mathcal{L}}X,Y) \eqno(5.3.2)$$
This allows (5.2.7) to be written in the form
$$J_P(t)=(I(t)F(0),F(0))/(P(0),P(0)),$$
$$   J_Q(t)=(1/M)(I(t)F(0),(1-{\mathcal{P}})P)/(Q(0),Q(0)) =0,\eqno(5.3.3)$$
where   
$$I(t)=\exp\{t(1-{\mathcal{P}}){\mathcal{L}}\},\eqno(5.3.4)$$
and
$$F(0)=(1-{\mathcal{P}}){\mathcal{L}}P(0)=(1-{\mathcal{P}})(-\Omega^2Q(0)+\alpha K(0))$$
$$=\alpha(1-{\mathcal{P}})K(0)=\alpha K(0)-\alpha^2( c_{KQ}/c_{QQ})Q(0)\eqno(5.3.5)$$ where we define the functions of temperture
$$c_{QQ}=(Q(0),Q(0)),\,\,\,\alpha c_{KQ}=(K(0),Q(0)).\eqno(5.3.6)$$
There is now (using $\rho$) a nonzero contribution from (5.2.6)
$$\Lambda_{KQ}=\alpha c_{KQ}/c_{QQ},\,\,\,\Lambda_{KP}=0. \eqno(5.3.7)$$
the latter being a consequence of time-reversal invariance (and the symmetry of the scalar product).

Note that here, because we have chosen the equilibrium distribution $\rho$ and therefore can use
(5.3.2),  $J_Q=0$ and $J_P\neq 0$; whereas in the approximation used in (5.2.13) (having 
chosen $\rho_0$) it was the other way round.

 For any $X$, (using (5.3.2))
$${\mathcal{P}\mathcal{L}}X=- {(X,{\mathcal{L}}Q(0))\over (Q(0),Q(0))}Q(0)-{(X,{\mathcal{L}}P(0))\over (P(0),P(0))}P(0), \eqno(5.3.8)$$
and if (5.3.1) is equivalent to
$${\mathcal{L}}Q=O(\Omega) \times Q,\,\,\,\,{\mathcal{L}}P=O(\Omega)\times P \eqno(5.3.9)$$
it follows that (5.3.8) is $O(\Omega)$; and then it is argued that we may approximate (5.3.4)
by 
$$I(t)\simeq \exp\{t{\mathcal{L}}\}. \eqno (5.3.10)$$
In this approximation, 
$$F(t)\simeq\exp(t{\mathcal{L}})F(0)=\alpha K(t)-\alpha^2(c_{KQ}/c_{QQ})Q(t)\eqno(5.3.11)$$
and
$$c_{PP}J_P(t)\simeq \alpha^2(K(t),K(0))-\alpha^3 (c_{KQ}/c_{QQ}) \{(K(t),Q(0))
+(Q(t),K(0))\}$$
$$+\alpha^4(c_{KQ}/c_{QQ})^2(Q(t),Q(0)). \eqno(5.3.12)$$
The last term in (5.3.11) (but not in (5.3.12)) is cancelled by (5.3.7).

The approximations (5.3.11) and (5.3.12) are to be inserted into (5.2.2) to get the hoped for Langevin equation. We wil now make some remarks about this `slow' approximation.

(a) Because of the dependence on $Q(t)$ on the right hand sides of (5.3.11) and (5.3.12),
the resulting equation has a different form from an ordinary Langevin equation.

(b) If we retain the term $\alpha K(t)$ on the right of (5.3.11) but discard everything else in (5.3.11)
and (5.3.12), the resulting equation (5.2.3) would reduce to just the original Heisenberg equation
of motion (3.1), not a Langevin equation at all.

(c)  Since ${\mathcal{L}}P=-\Omega^2 Q+\alpha K$,
the assumption (5.3.9) seems to require that $\alpha$ as well as $\Omega$  be small.

(d) We can test the `slow' approximation in the special case when $K$ is a linear function
$$K=\sum c_iq_i ,\eqno(5.3.13)$$
 when the ordinary Langevin equation is well known and is in fact exact.
This equation amy be written in the form (5.2.3) with 
$$J_Q=0,\,\,\, J_P(t)=\alpha^2(1/M)\sum(c_i/m_i\omega_i)^2\cos(\omega_it)=\alpha^2(MT)^{-1}\mathrm{tr}\{K_0(t)K_0(0)\rho_E\},$$
$$F(t)=\alpha K_0(t)-\alpha^2 Q(0)\sum (c_i/m_i\omega_i)^2\cos(\omega_it),\eqno(5.3.14)$$
where we have used the notation of (2.6).

Let us see if we can get (5.3.14) from (5.3.11) and (5.3.12) in any approximation. The nearest we can get
is to write (5.3.11) and (5.3.12) as
$$F(t)=\alpha K_0(T)+O(\alpha^2),$$
$$J_P(t)=\alpha^2(1/MT)\langle K_0(t)K_0(0)\rangle_0+O(\alpha^3).\eqno(5.3.15)$$
This does reproduce (5.3.14) \textit{except} for the $Q(0)$ term at the end of (5.3.14),
which has to be considered as being of the neglected orders.

This example confirms our expectation in (c) above, that $\alpha$ as well as $\Omega$
has to be regarded as small. But even then the `slow' approximation seems to be
incapable of reproducing the whole of the correct result (5.3.14) including the $Q(0)$ term.

It is worth remarking that the Langevin equation for linear model  (5.3.13) can alternatively be written
in the form (5.2.3) with $J_P=0$ and $J_Q\neq 0$, and then there is no $Q(0$  term; but the slow approximation to the generalized Langevin equation forces the alternative form with $J_Q=0$.

Finally, we note that if we are prepared to neglect higher powers of $\alpha$ on the right
hand sides, we can immediately make Senitzky's approximation (3.5) for example,
without going via the generalized Langevin equation.

We conclude that the `slow' approximation to the generalized Langevin equation is not straightforward.

\section{Conclusions}

Although the Langevin equation (for a simple system interacting with a large environment) is often used, not so much attention has been given to
judging its validity. We have critically examined two possible methods for doing this, with
particular attention to the quantum case.
 
 The first is a direct approach due to Senitzky \cite{senitzky}. He made approximations
 whose validity is not obvious in general. We identify one region in which the
 method may be justified, and that is where the measured frequency is close to the
 natural frequency of the free system. This condition cannot be formulated when the system
 is nonlinear, although there are important examples where this is the case.
 
 The second approach (which also requires the system to be a linear one) seems to be completely different (see for example \cite{zwanzig}). It proceeds by using a projection operator
 to  a generalized
 Langevin equation, which is an exact consequence of the Heisenberg equations of motion,
 but which is of little practical use as it stands. In section 5.2, we argued that the identification of
 the noise and dissipation terms is ambiguous in the generalized equation.
 
  Then we studied the assumption that the system variables
 are `slow' compared to the environment ones. This leads to an equation looking like
 an ordinary Langevin equation. We examine the steps going into this derivation,
 particularly by comparison with the model in which all the equations of motion are linear,
 for which the Langevin equation is easily established and is exact. We argue that
 the `slow' approximation necessarily entails neglect also of terms
 of higher order in the coupling strength $\alpha$, as well as in the frequency ratio.
 But even then the correct Langevin equation requires a selective choice of orders of
 $\alpha$.

JF would like to thank CNPq, Brazil, for a grant.

\end{document}